\documentclass[a4paper,11pt]{article}
\usepackage{pos}
\usepackage{wrapfig}

\title{Giant Radio Array for Neutrino Detection (GRAND)}
\ShortTitle{GRAND}

\author*[a]{Sijbrand de Jong}
\author[b]{the GRAND Collaboration}

\affiliation[a]{Radboud University Nijmegen and Nikhef,
  Heyendaalseweg 135, Nijmegen, The Netherlands.}
\affiliation[b]{https://grand.cnrs.fr}

\emailAdd{sijbrand@hef.ru.nl}

\abstract{
GRAND is a newly proposed series of radio arrays with a combined area of
200,000~km$^2$, to be deployed in mountainous areas. Its primary goal is to
measure cosmic ultra-high-energy tau-neutrinos ($E>1$~EeV), through the
interaction of these neutrinos in rock and the decay of the tau-lepton in
the atmosphere. This decay creates an air shower, whose properties can
be inferred from the radio signal it creates. The huge area of GRAND makes
it the most sensitive instrument proposed to date, ensured to measure
neutrinos in all reasonable models of cosmic ray production and propagation.
At the same time, GRAND will be a very versatile observatory with
enormous exposure to ultra-high-energy cosmic rays and photons.

This talk covers the scientific motivation, as well as the staged
approach required in the R\&D stages to get to a final design that will make
the construction, deployment and operation of this vast detector affordable.}

\FullConference{%
  40th International Conference on High Energy physics - ICHEP2020\\
  July 28 - August 6, 2020\\
  Prague, Czech Republic (virtual meeting)
}

\begin{document}
\maketitle

\begin{wrapfigure}{r}{8.8cm}
\vspace*{-5mm}
\includegraphics[width=8.8cm]{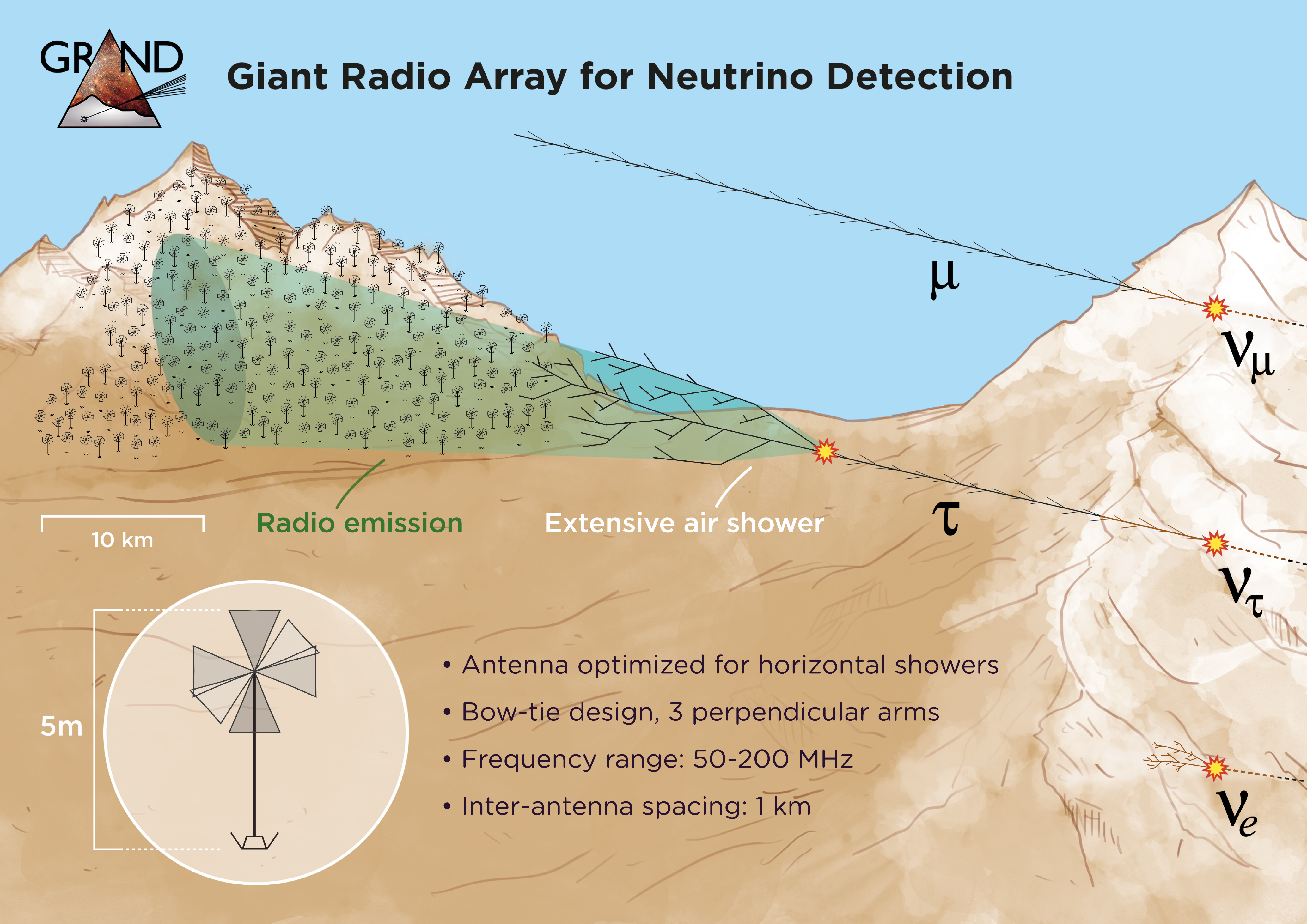}\\[-8mm]
\caption[GRAND overview]{\label{fig:overview}
                                    \em
                                    Overview of the GRAND ultra-high-energy neutrino detection scheme.
                                    }
\vspace*{-5mm}
\end{wrapfigure}
\section{Introduction}
\vspace*{-4mm}
GRAND is a planned Giant Radio Array for Neutrino Detection~\cite{bib:GRANDwhitepaper}.
The principle of the GRAND observatory layout is shown in Fig.~\ref{fig:overview}.
It should be large enough to detect ultra-high-energy cosmic neutrinos, with an energy above an EeV or so,
for all standard theory scenarios. This turns out to require a 200\,000 km$^2$ array.
This is huge and too big to be realised in one go today. Therefore a staged approach is followed, where at present a 300 antenna GRAND Proto300 array is being realised and a more than 10\,000~km$^2$ array would be the
logical next step.

\vspace*{-4mm}
\section{Ultra-high-energy neutrino production and detection}
\vspace*{-4mm}
Ultra-high-energy cosmic neutrinos are assumed from two sources.
They can come directly from the sources of ultra-high-energy (UHE) cosmic rays,
where in the interactions of cosmic rays also pions and thereby neutrinos are being produced.
IceCube probably detected such neutrinos in the PeV energy range~\cite{bib:IceCubeBlazar}.
Or these neutrinos can come from interactions of UHE cosmic rays with
the cosmic photon background from CMB and other sources,
where also pions can be produced, thus creating neutrinos.
The neutrinos that are produced are of the muon and electron type and
oscillation to tau neutrinos will cause as many tau neutrinos as any of the other two types to reach the Earth.
The yield of cosmogenic UHE neutrinos can be estimated knowing the UHE cosmic ray spectrum,
but e.g. the unknown composition, source distribution, etc. give appreciable uncertainty.
Extrapolating the IceCube spectrum~\cite{bib:IceCubespectrum}
may give an indication of the expected UHE neutrino yield from sources,
but is rather uncertain due to the galactic to extra-galactic transition in the spectrum.

The detection principle is that UHE neutrinos develop a sizable cross section with ordinary matter.
Therefore at energies in the EeV region there is a sizable probability that a neutrino interacts
in a few kilometres of Earth.
At these energies neutrinos basically never penetrate the full Earth and it is no longer the
volume of the target that determines the yield, but the surface with a few km thickness.
The strategy then is to look for neutrinos that interact with mountains or are Earth skimming.
On a charged current interaction either an electron, muon or tau emerges.
On the right hand side of Fig.~\ref{fig:overview} the interactions for each type of neutrino with
a mountain are illustrated.
Electrons will quickly get stuck, producing an electromagnetic shower,
with a small probability of a signal escaping the Earth.
Muons will escape, but do not leave much signal density in the Earth or atmosphere
and are therefore hard to detect.
Taus on the other hand, due to their limited, but finite lifetime
will mostly leave the mountain and decay in the atmosphere in about 50~km or so, where an air shower develops.
The aim is to detect these air showers using their radio signal.
Since the atmosphere is transparent for this radio frequency radiation,
the shower can be observed also at many km after it died out.
The radio detection has to be optimised for horizontal showers, which is a challenge,
but has been demonstrated as possible in principle~\cite{bib:AERAhas}.

\section{GRAND as multi-messenger observatory}
\begin{figure}[t]
\vspace*{-5mm}
\hfill
\includegraphics[height=5cm]{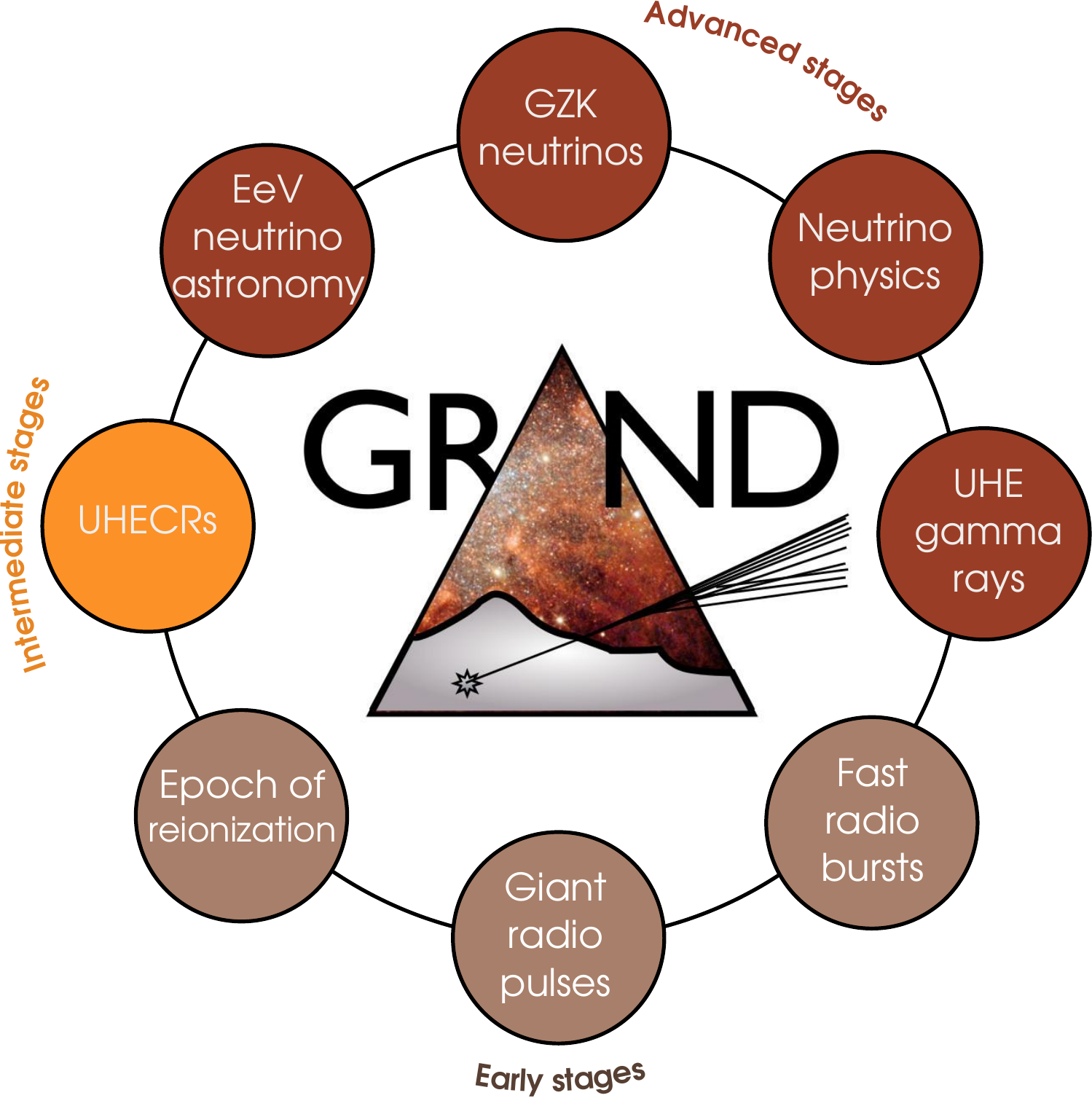}\hfill
\includegraphics[height=5cm]{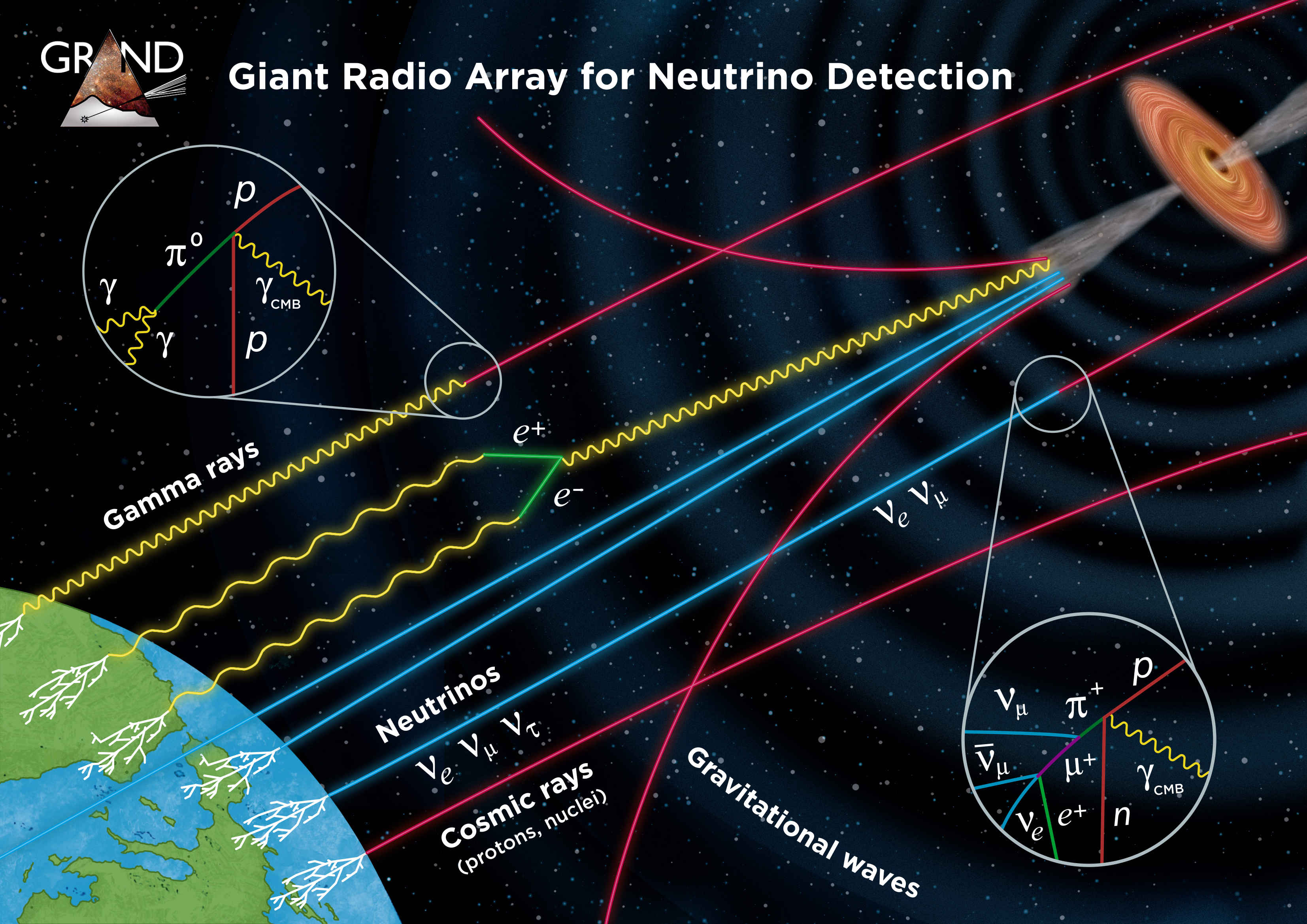}\hspace*{10mm}\\[-8mm]
\caption[GRAND multi-messenger]{\label{fig:multimessenger}
                                    \em
                                    GRAND as versatile observatory and its multi-messenger capabilities.
                                    The physics and astronomy topics that will be covered are listed in the diagram
                                    on the left. The sources and propagation of the various messengers that can be
                                    detected by GRAND are illustrated in the picture on the right hand side.
                                    }
\vspace*{-4mm}
\end{figure}

\vspace*{-4mm}
Aside from the primary goal of UHE neutrino detection,
for both particle physics and astrophysics measurements,
a huge radio detector will also cover many other science goals,
such as the detection of UHE photons and cosmic rays, making it a true multi-messenger observatory.
This is illustrated on the right hand side in Fig.~\ref{fig:multimessenger}
Of course this also gives a background to the neutrinos, but the distinction is not so difficult,
e.g. using the direction, where no photons or cosmic rays are expected from lines of sight with many km of Earth.
GRAND will also offer opportunities for the detection and measurement of fast radio bursts and giant radio pulses,
as well as a measurement of the epoch of re-ionisation, as illustrated in the diagram on the left
in Fig.~\ref{fig:multimessenger}, but this will not be covered in this presentation.

\vspace*{-4mm}
\section{GRAND predicted performance}
\begin{figure}[b]
\vspace*{-5mm}
\includegraphics[height=4.3cm]{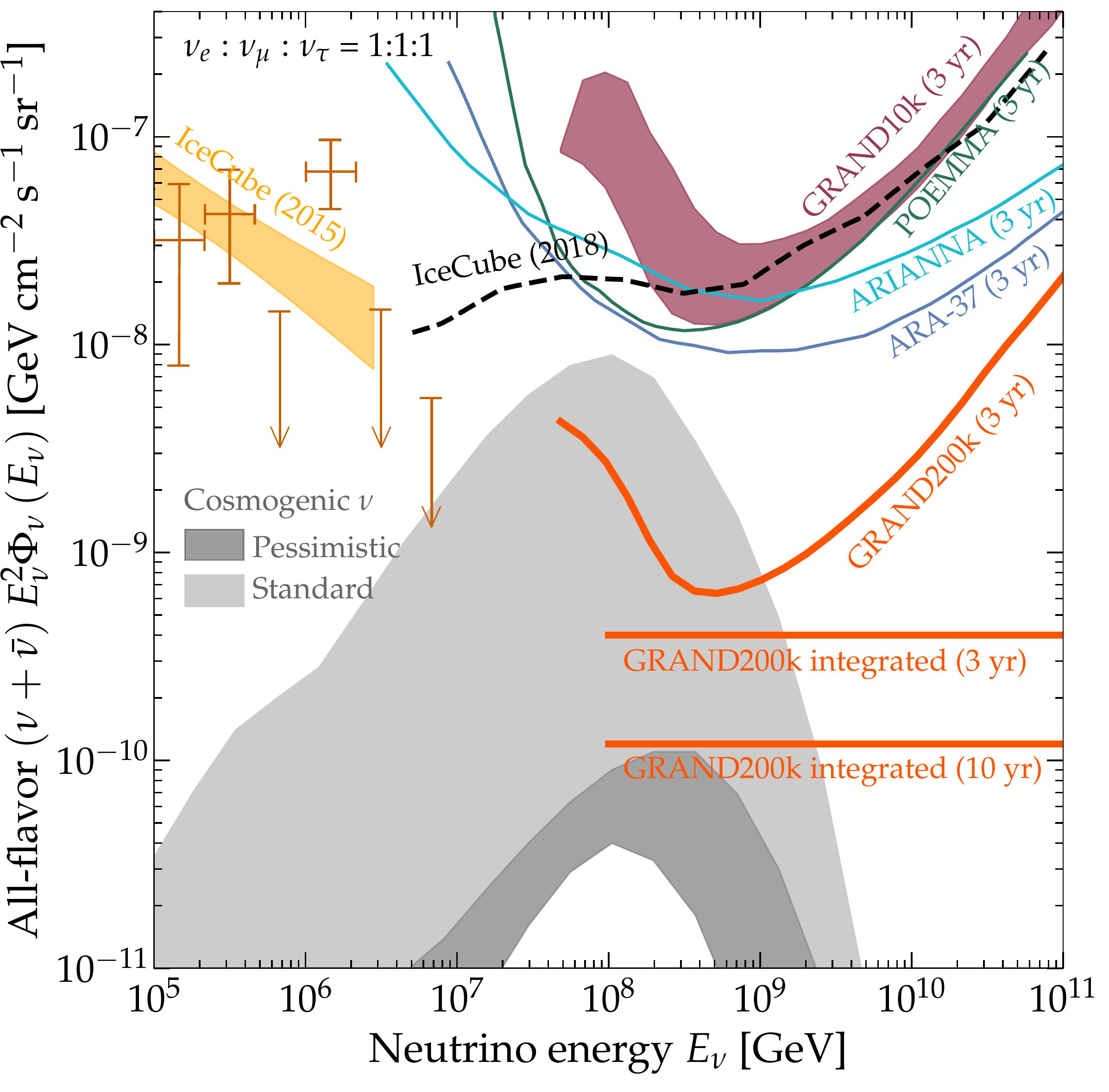}\hfill
\includegraphics[height=4.3cm]{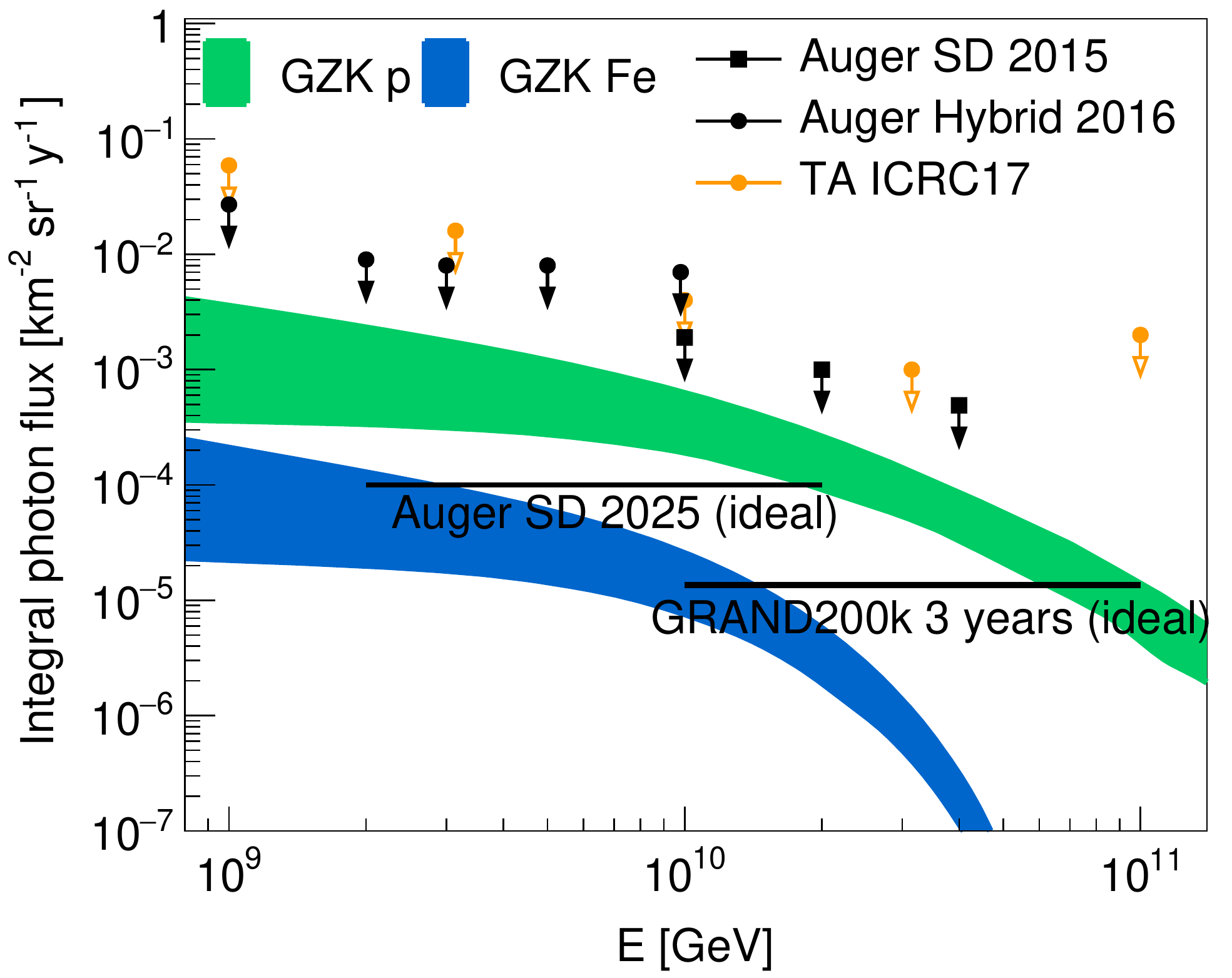}\hfill
\includegraphics[height=4.3cm]{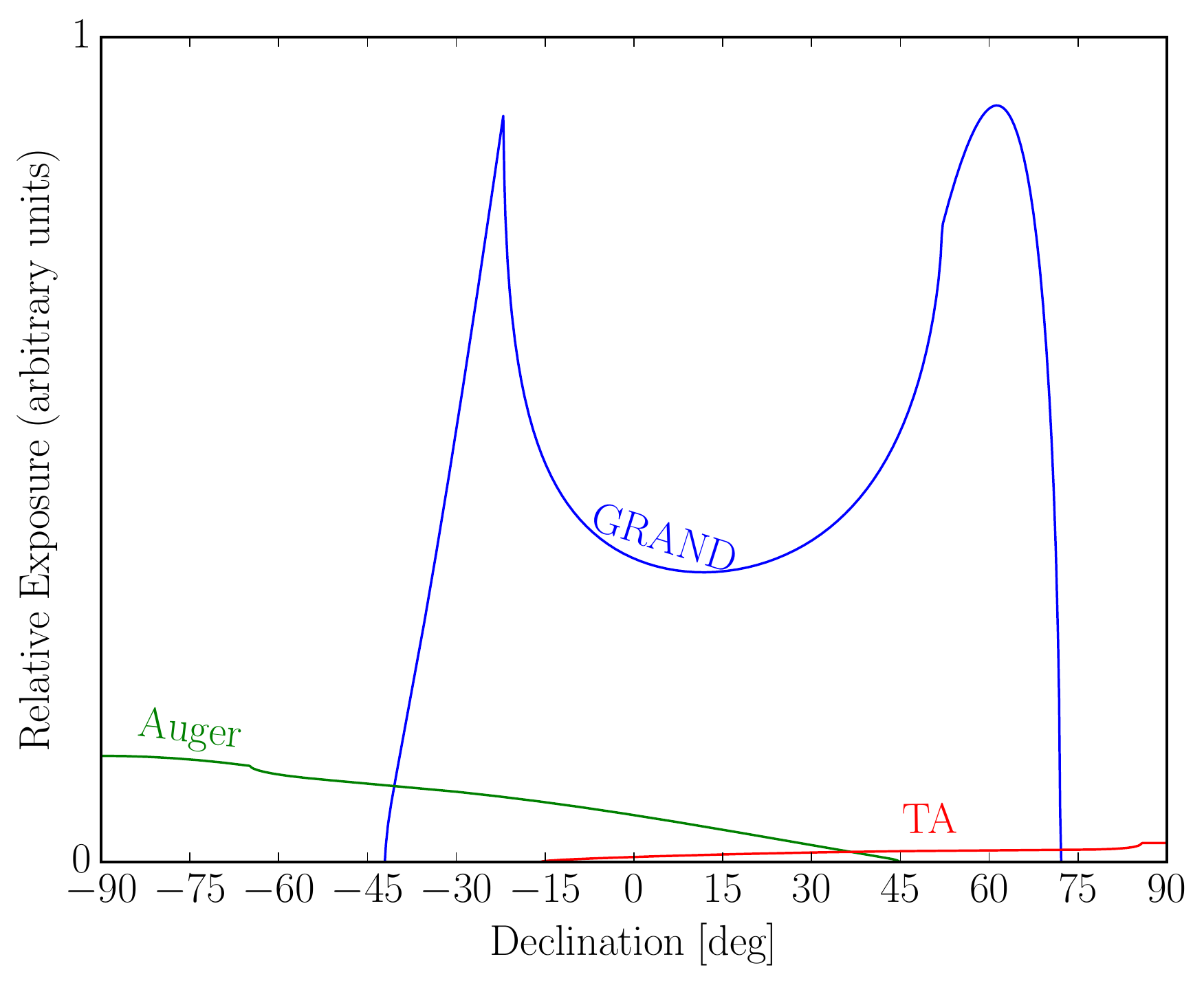}\\[-8mm]
\caption[GRAND multi-messenger]{\label{fig:fluxes}
                                    \em
                                    Estimated flux sensitivities for GRAND and limits and sensitivity
                                    of other present and future observatories for UHE neutrinos on the left
                                    and UHE photons in the middle.
                                    On the right the relative exposure for UHE cosmic rays for GRAND
                                    compared to those of the Pierre Auger and Telescope Array Observatories.
                                    Figures from~\cite{bib:GRANDwhitepaper} with references to data and predictions
                                    mentioned therein.
                                    }
\end{figure}
\vspace*{-4mm}
An end-to-end simulation and reconstruction pipeline has been set up, with
the development of efficient semi-analytical methods for the modelling of
radio emissions~\cite{bib:emissionmodel}.
Dedicated
reconstruction methods for very inclined air-showers are also being
developed, with promising results in terms of angular resolution ($\sim 0.1^{\circ}$)
and on the composition estimator~\cite{bib:thesisValentin}.
From this, the required size of the array with a 1 km detector station spacing has been determined to be about
200\,000 km$^2$ for a certain detection of UHE cosmogenic neutrinos over the full range of
the standard theoretical models~\cite{bib:GRANDwhitepaper}.
This is illustrated in Fig.~\ref{fig:fluxes} on the left,
where the predicted GRAND performance is plotted with several existing limits and sensitivity of other
future observatories.
In case of no detection, something extraordinary is going on and we would have to completely rethink our view
of the UHE particle universe, this would be the scientific jack pot.
From the simulation also the sensitivity to UHE photons is illustrated in the middle plot of Fig.~\ref{fig:fluxes},
where also for UHE photons detection is expected for most reasonable scenarios.
GRAND will also collect a very large sample of UHE cosmic rays, which is illustrated on
the right hand side of Fig.~\ref{fig:fluxes}, where more than an order of magnitude more events are
expected to be detected by GRAND than are recorded with the present Pierre Auger and Telescope Array
Observatories.

\section{GRAND sites}
\begin{wrapfigure}{r}{6.1cm}
\vspace*{-22mm}
\centerline{\hspace*{-4mm} \includegraphics[width=6.4cm]{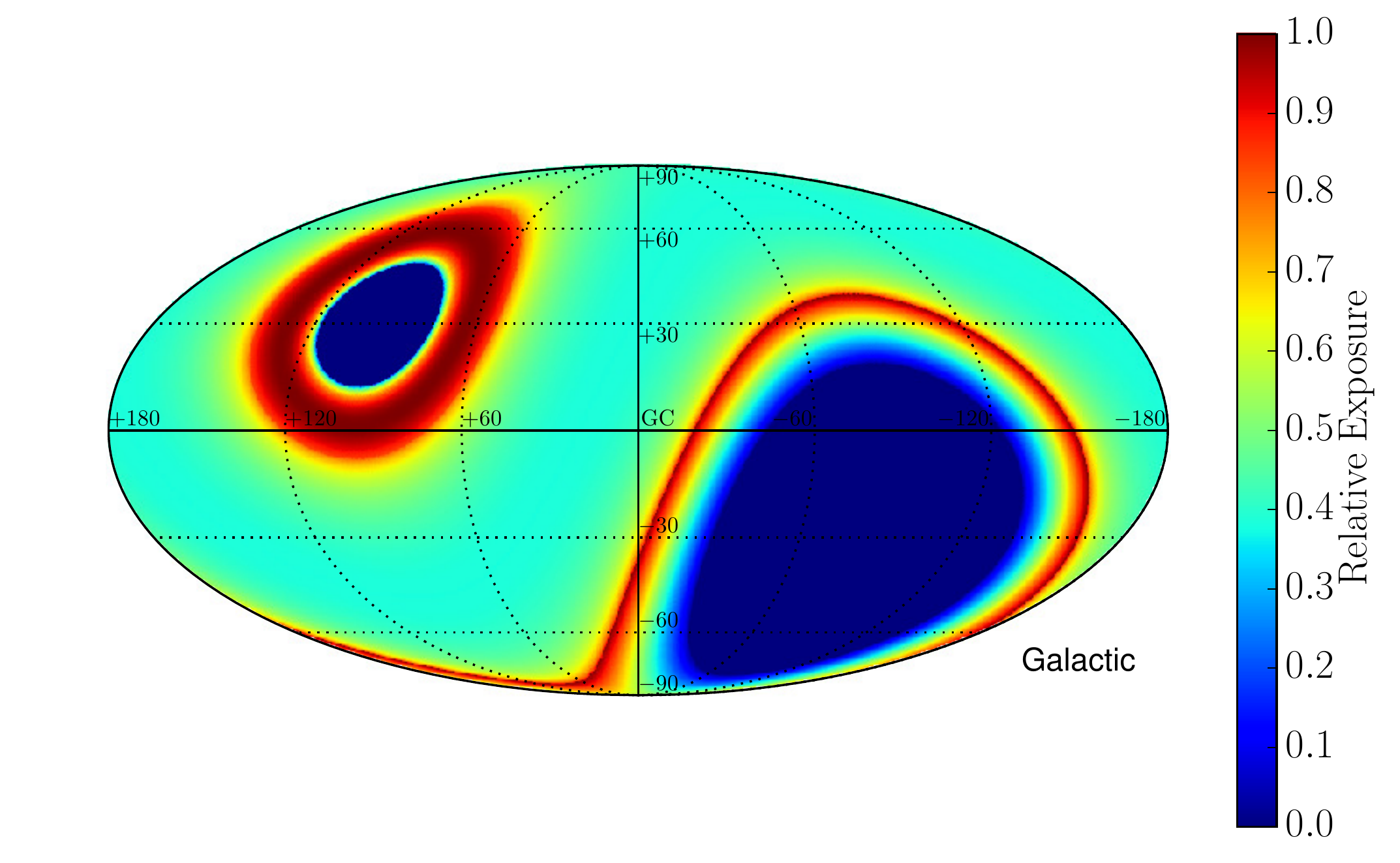}}
\caption[GRAND one site exposure map]{\label{fig:exposuremap}
                                   \em
                                   Exposure map for 1 GRAND site.
                                   }
\vspace*{-5mm}
\end{wrapfigure}
\vspace*{-4mm}
But there is also astrophysics reason to distribute this area in a number of sites.
Due to the sensitivity to horizontal showers, at any one time a circular wedge of the universe is viewed.
Although from one site, integrated over time,
Fig.~\ref{fig:exposuremap},
covering all or most of the sky at any time will
give much more opportunity for observation of transients about 80\% of the sky is observed as shown in
 that are also observed by other instruments.
Therefore, to be able to optimally contribute to multi-messenger astronomy, it makes sense to have
of the order of ten detector sites distributed around the world.
One of such sites should therefore be about 10\,000 to 20\,000 km$^2$
and on its own has already a large enough exposure to deliver a very exciting science programme
from the moment the first site starts collecting data, while the other sites are still being realised.
A detector array of 200\,000 km$^2$, three times the size of the Czech Republic,
will be very hard to realise in one location for reasons of available suitable land surface area,
technical realisation and maintenance, and for political reasons.\\[-5mm]

\section{GRAND roadmap}
\begin{wrapfigure}{r}{10cm}
\vspace*{-15mm}
\includegraphics[width=10cm]{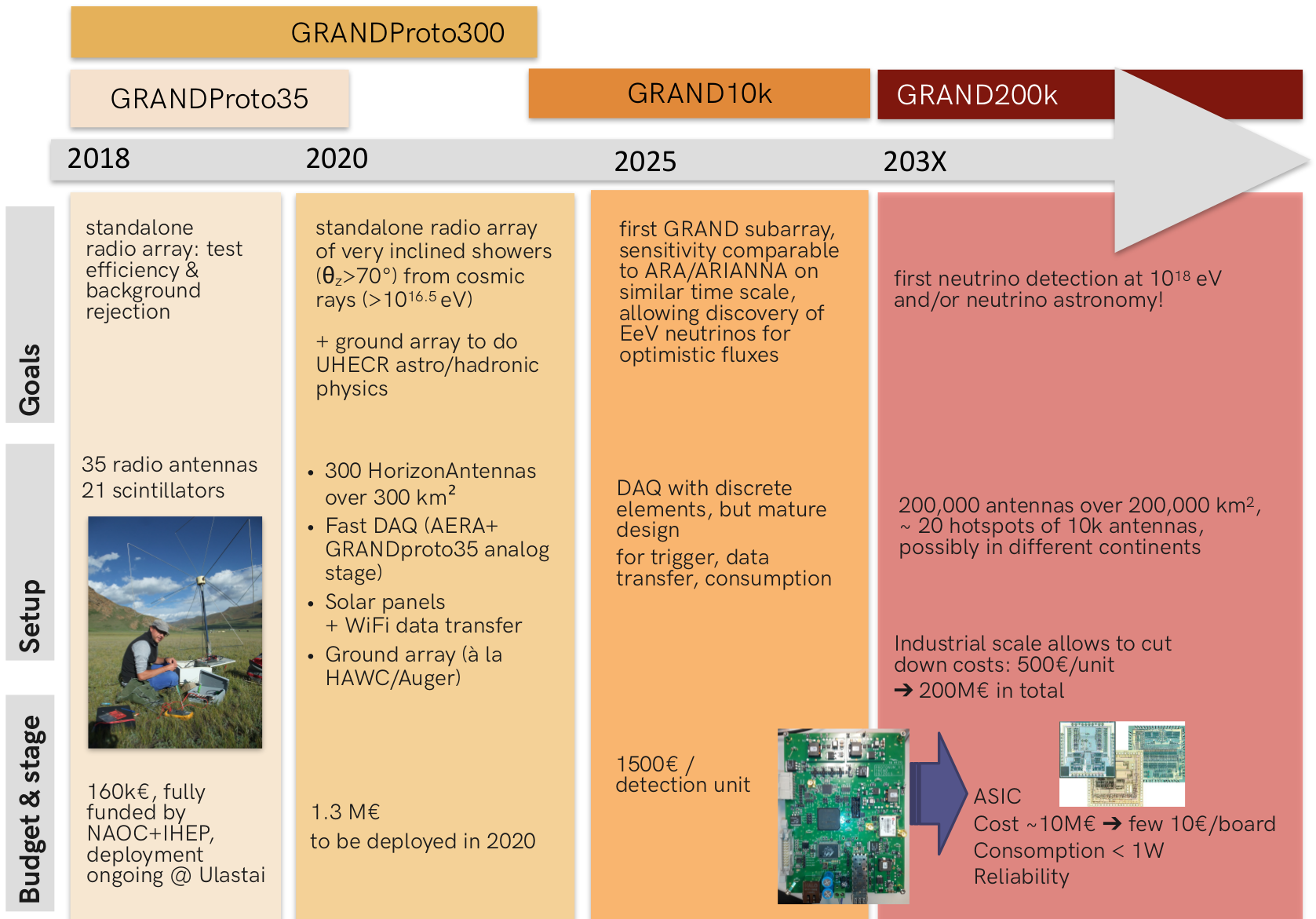}\\[-8mm]
\caption[GRAND multi-messenger]{\label{fig:roadmap}
                                    \em
                                    GRAND road map.
                                    }
\vspace*{-5mm}
\end{wrapfigure}
\vspace*{-4mm}
The staged approach allows the ultimate goal of 200\,000 km$^2$ to be attained in steps
and leads to a staged approach, where one 10\,000--20\,000 km$^2$ array can be deployed after the other.
For such a huge array to be feasible, the detection not only needs to work well,
which has been demonstrated~\cite{bib:AERAtalkICHEP2020},
but it also needs to be financially viable, easy to deploy and extremely reliable.
In addition, it needs to provide its own trigger,
which has been demonstrated to be possible~\cite{bib:trigger},
but for which the purity needs to be improved.
For these reasons the GRAND roadmap in Fig.~\ref{fig:roadmap}
starts with a series of smaller scale prototypes.
As a first step GRANDProto35 has demonstrated the principle on a small array of
35 detector stations in on a site in China with the specification compatible with a large array.
Currently, GRANDProto300, a 300 detector station array on 300~km$^2$, is being produced,
where for the first time in this roadmap steps are taken towards making the detector stations
cheaper, easier to deploy and more reliable.
GRANDProto300 is also a main vehicle to research self-triggering strategies.

\section{GRANDProto300 progress}
\begin{wrapfigure}{r}{4.8cm}
\vspace*{-15mm}
\centerline{\includegraphics[width=4.8cm]{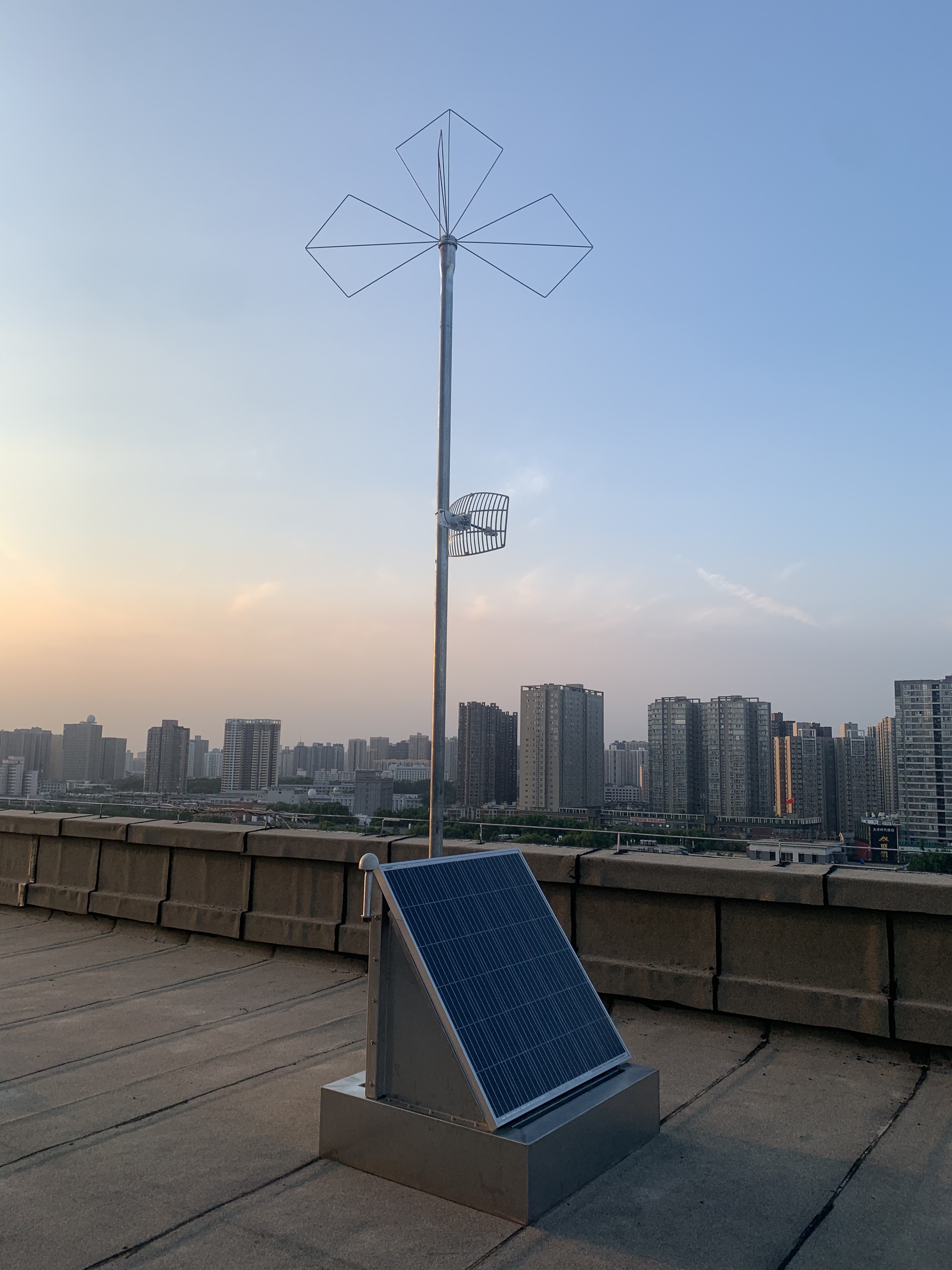}}
\caption[GRAND one site exposure map]{\label{fig:Proto300station}
                                   \em
                                   Prototype station for the GRANDProto300 detector station.
                                   }
\vspace*{2mm}
\centerline{\includegraphics[width=4.8cm]{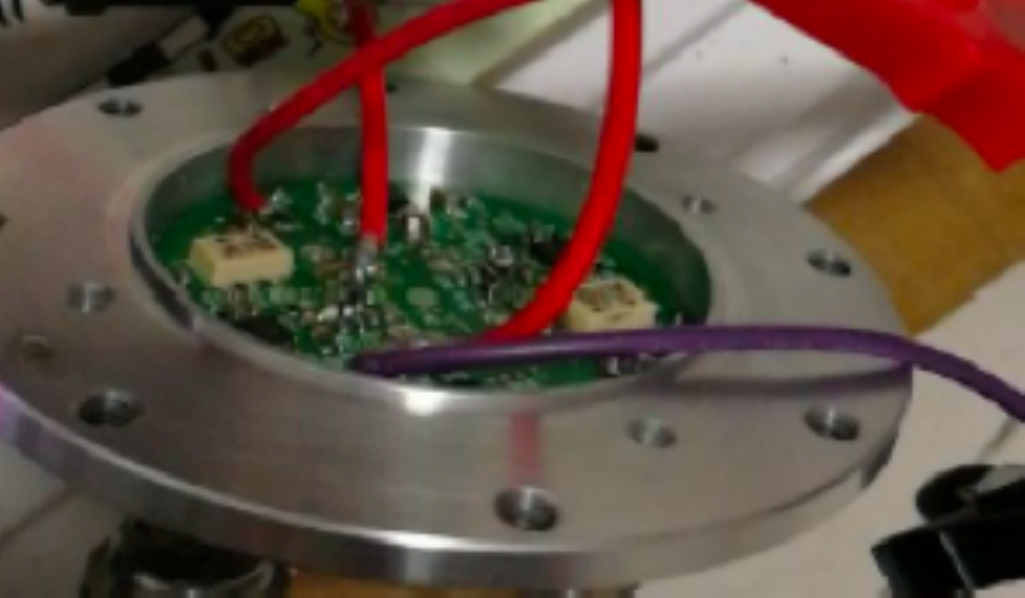}}
\caption[LNA]{\label{fig:LNA}
                                   \em
                                   Low Noise front-end Amplifier prototype.
                                   }
\vspace*{-6mm}
\end{wrapfigure}
\vspace*{-4mm}
The radio detection station in GRANDProto300, a prototype shown in Fig.~\ref{fig:Proto300station}
 are autonomously operated.
They have to create their own power and communicate through a wireless connection.
The station consists of the receiver antenna mounted on a 3.5~m pole,
where at half height also the wireless communication antenna is mounted.
The pole is supported by a box that holds the solar panel for power harvesting and
inside a battery to store and distribute power for 24/7 functioning
and the electronics for digitising, analysing, triggering and communicating to the central DAQ system.
Underneath the box is an area that can be filled with sand and rocks on location to mechanically
stabilise the structure, also in strong wind.
The design is optimised for maintenance and modification of the set-up at the cost of the price.
The final design needs to be even much cheaper and more robust, but should then not need servicing.

The detection frequency range of choice is 50-200 MHz.
A significant amount of work has been invested into an antenna design that is optimised
for the detection of horizontal showers.
The current  design uses a butterfly shape in three polarisation direction:
Two two-arm antennas for the horizontal  
and a single-arm for the vertical polarisation.

The arms are relative small for the frequency range,
which\\[-5mm]

\noindent
\begin{wrapfigure}{l}{5.5cm}
\vspace*{-4mm}
\centerline{\includegraphics[width=5.5cm]{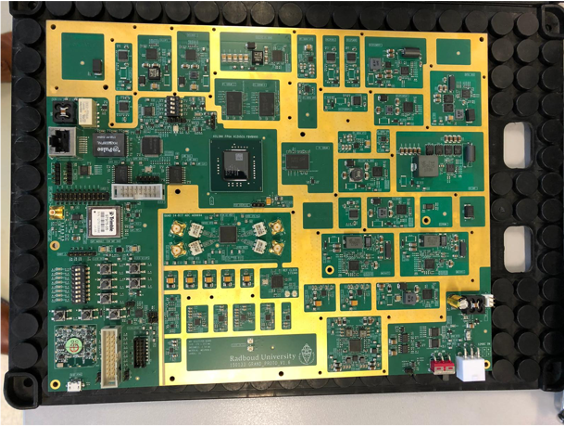}}
\caption[Digitiser]{\label{fig:digitiser}
                                   \em
                                   GRANDProto300 digitiser.
                                   }
\vspace*{-5mm}
\end{wrapfigure}
was necessary to get a good impedance match to the low noise amplifier
over the full frequency range and it also helps for mechanical stability.
Combined with the long pole on which the antenna is mounted it
reduces ground reflections for horizontal showers and the influence of
the rest of the mechanical structure at the foot of the pole.

The electronics consists of a low noise amplifier, see Fig.~\ref{fig:LNA},
AC connected directly to the antenna at the top of the pole.
In the electronics box the signal is frequency filtered, amplified and 
digitised with 500 Mega samples of 14 bits per second, see Fig.~\ref{fig:digitiser} for the digitiser board.
An integrated FPGA and CPU chip analyses the signal, provides a station trigger and gathers
housekeeping information, e.g.\ for calibration.
Prototype versions are in hand and are being tested (Figs.~~\ref{fig:LNA} and~\ref{fig:digitiser}).
The final board is ready for production.
Station trigger information is sent to a central DAQ station that combines the information in a level 2 trigger,
which determines if the station information is read by the central DAQ.
After the level 2 trigger, the data processing is centralised.

\vspace*{-4mm}
\section{Outlook} 
\vspace*{-4mm}
GRAND is a huge detector that has been designed to have a no-loose theorem for UHE neutrino
observation, if they cannot be observed by GRAND, some new (astro)physics must be going on. 
But GRAND is a much more versatile next generation tool, also being able
to detect UHE photons and cosmic rays and capable of doing exciting radio astronomy.
GRAND is to be realised in a step-by-step approach and a road map towards its construction
has been established. The next step on this road map is the construction of a 300 detector
station prototype, GRANDProto300, which is well underway. In this step the focus is on the
self-triggering, which can be developed and tested with cosmic ray events.
The step after GRANDProto300 still poses large challenges, such as making the detector
stations much cheaper, much easier to deploy and much more robust, all at the same time.
The GRAND collaboration is making good steps forward, but will need to grow considerably
to be able to achieve its goals. Therefore, more help from new collaborators looking for
exciting challenges and a grand goal is much welcomed.

\end{document}